\documentclass[aps,arxiv,twocolumn,superscriptaddress]{revtex4-1}  

\usepackage[utf8]{inputenc}
\usepackage[T1]{fontenc}

\usepackage{graphicx}
\usepackage{bm}
\usepackage{amsmath}
\usepackage{amssymb}
\usepackage{mathtools}
\usepackage[version=4]{mhchem}
\usepackage{microtype}

\usepackage[colorlinks, allcolors=blue]{hyperref}
\usepackage{xcolor}
\hypersetup{pdftitle={Li$_{x}$Si phase diagram},
  pdfauthor={Nongnuch~Artrith}}
\makeatletter
\let\Hy@backout\@gobble
\makeatother

\begin{document}


\title{Constructing first-principles phase diagrams of amorphous
  \ce{Li_xSi} using machine-learning-assisted sampling with an
  evolutionary algorithm}

\author{Nongnuch Artrith}
\email{nartrith@berkeley.edu}
\affiliation{%
  Department of Materials Science and Engineering,
  University of California, Berkeley, CA, USA}
\affiliation{%
  Materials Science Division, Lawrence Berkeley National
  Laboratory, Berkeley, CA, USA}
\author{Alexander Urban}
\affiliation{%
  Department of Materials Science and Engineering,
  University of California, Berkeley, CA, USA}
\affiliation{%
  Materials Science Division, Lawrence Berkeley National
  Laboratory, Berkeley, CA, USA}
\author{Gerbrand Ceder}
\email{gceder@berkeley.edu}
\affiliation{%
  Department of Materials Science and Engineering,
  University of California, Berkeley, CA, USA}
\affiliation{%
  Materials Science Division, Lawrence Berkeley National
  Laboratory, Berkeley, CA, USA}
\date{\today}

\begin{abstract}
  The atomistic modeling of amorphous materials requires structure sizes
  and sampling statistics that are challenging to achieve with
  first-principles methods.
  Here, we propose a methodology to speed up the sampling of amorphous
  and disordered materials using a combination of a genetic algorithm
  and a \emph{specialized} machine-learning potential based on
  artificial neural networks (ANN).
  We show for the example of the amorphous LiSi alloy that around 1,000
  first-principles calculations are sufficient for the ANN-potential
  assisted sampling of low-energy atomic configurations in the entire
  amorphous \ce{Li_xSi} phase space.
  The obtained phase diagram is validated by comparison with the results
  from an extensive sampling of \ce{Li_xSi} configurations using
  molecular dynamics simulations and a \emph{general} ANN potential
  trained to $\sim$45,000~first-principles calculations.
  This demonstrates the utility of the approach for the first-principles
  modeling of amorphous materials.
\end{abstract}

\maketitle


\section{Introduction}
\label{sec:introduction}

Amorphous phases determine the properties of many technologically
relevant materials, such as catalysts for water oxidation and
reduction,\cite{jacs136-2014-2843, nm15-2015-4511, nc5-2014-4059}
lithium and sodium ion battery electrodes,\cite{am25-2013-4966,
  ac125-2013-4731} electrodes for solid oxide fuel
cells,\cite{aem5-2015-1400747, as-2017-1700337} and the solid
electrolyte interphase at the electrode/electrolyte
boundary.\cite{nl15-2015-2011}
Owing to the absence of long-range order, amorphous structures cannot be
characterized by conventional diffraction techniques.
Other characterization methods, such as X-ray or neutron pair
distribution function measurements, only provide indirect structural
information.
First principles density-functional theory (DFT)~\cite{pr136-64-846,
  pr140-65-a1133, jcp136-2012-150901} offers predictive insight into
phase stabilities and structures~\cite{ncm2-2016-16002} and is an
attractive alternative for the characterization of atomic structures and
their properties.
However, the high computational cost of DFT calculations confines the
method to structure models containing at most a few hundred atoms and
moderate sampling, a limitation that in practice makes it challenging to
model many amorphous or disordered phases.

To overcome these sampling and size limitations, recently a number of
machine-learning techniques have been proposed that can be used to
efficiently interpolate first principles energies and atomic
forces.\cite{prl98-2007-146401, prl104-2010-136403, prl108-2012-058301,
  prl117-2016-135502, pnas113-2016-8368, nc8-2017-872}
Machine-learning potentials (MLPs) have enabled large-scale simulations
of complex materials such as nanoalloys,~\cite{nl14-2014-2670,
  cms110-2015-20} metal-oxide nanoparticles,~\cite{ac6-2016-1675} and
amorphous carbon~\cite{prb95-2017-94203} with an accuracy that is close
to the reference method, but at a fraction of the computational cost.
However, to reliably sample all relevant atomic interactions MLPs have
to be trained on extensive databases of first-principles calculations.
Depending on the number of chemical species, several thousand to tens of
thousands of reference data points might be required to obtain a
reliable MLP.
By their nature, amorphous phases exhibit a large variety of local
structural motifs, making the compilation of complete structural
reference databases even more challenging.

Here we propose an alternative approach for the calculation of
first-principles phase diagrams of amorphous materials with
\emph{machine-learning-assisted} sampling.
Instead of constructing a \emph{general} MLP based on tens of thousands
of reference calculations, we employ a \emph{specialized} MLP trained on
only around one thousand DFT calculations of small structures, and which is
used specifically for the sampling of near-ground-state structures with
a genetic algorithm.
The relevant structures identified by this sampling approach are
recomputed using DFT to obtain an accurate first-principles phase
diagram.
Finally, we validate the approximate sampling approach by training a
fully converged \emph{general} MLP that makes possible extensive
molecular dynamics simulations.

We consider \ce{Li_xSi} alloys as a prototypical example of an
amorphous electrode material.
Nanostructured amorphous Si is a potential high-capacity negative
electrode material for Li-ion batteries.\cite{am25-2013-4966,
  aem4-2013-1300882, jmca1-2013-9566, ra6-2016-87778}
The lithiation of crystalline Si has previously been studied
with first-principles calculations on small model structures,
and through mesoscale simulation techniques.\cite{jes156-2009-454,
  cjp87-2009-625, jes157-2010-392, jpcm22-2010-415501,
  jpcc115-2011-2514, jpcl2-2011-3092, nl13-2013-2011,
  jacs134-2012-14362, nl14-2014-4065, jpcc119-2015-3447,
  pccp17-2015-3832}
In the present article, we consider significantly larger structure
models with more than 600~atoms, which provides additional insight into
the energetics of the amorphous LiSi phase.

The paper is structured as follows:
In section~\ref{sec:energy-models}, we detail all simulation parameters
and recap the artificial neural network potential method, as well as the
descriptor technique used to capture the local atomic environment.
This is followed by a description of the construction of the specialized
MLP and the genetic algorithm in section~\ref{sec:methods-GA} and the
molecular dynamics sampling in section~\ref{sec:MD-sampling}.
In the next section~\ref{sec:phase-space}, the phase space of amorphous
\ce{Li_xSi} structures that is sampled by the genetic algorithm is
compared with that of the molecular dynamics heat-quench simulations.
The results are assessed in a final discussion
section~\ref{sec:discussion}.

\section{Energy Models}
\label{sec:energy-models}

\subsection{Density-functional theory calculations}
\label{sec:DFT}

All density-functional theory (DFT) calculations were carried out using
the Vienna Ab-Initio Simulation Package (VASP).\cite{prb54-1996-11169,
  cms6-1996-15}
The exchange-correlation functional by Perdew, Burke, and Ernzerhof
(PBE)~\cite{prl77-96-3865, prl78-97-1396} and projector-augmented wave
(PAW) pseudopotentials~\cite{prb50-1994-17953} were used.
The energy cutoff of the plane wave basis set was generally 520~eV, and
a $k$-point density of 1000~divided by the number of atoms was used for
the Brillouin-zone integration following the recommendations from
reference~\citenum{cms50-2011-2295}.
VASP input files were generated using the Python Materials Genomics
tool (\emph{pymatgen}).\cite{cms68-2013-314}
In geometry optimizations, energies were generally converged to
0.05~meV/atom and the convergence thresholds for atomic forces was
50~meV/\AA{}.

\subsection{Machine-learning potentials with Chebyshev descriptor}
\label{sec:methods-MLP}

The idea behind machine-learning potentials (MLPs) is to interpolate
\emph{first principles} potential energies and atomic forces using
machine learning techniques, such as artificial neural networks
(ANNs)~\cite{prl98-2007-146401, prl108-2012-058301} or Gaussian process
regression.\cite{prl104-2010-136403}
For the structure space that it is trained on, an MLP can be nearly as
accurate as its reference method at a fraction of the computational
cost.\cite{acie56-2017-12828}
Several approaches in this spirit have been proposed, varying in the
details of the machine learning model, the atomic structure descriptor,
and the interpolated quantity (total energy vs.\ atomic
energy).\cite{ijqc115-2015-1051, prb92-2015-45131, pccp18-2016-13754,
  prl117-2016-135502}

In the present work, we employ ANNs to interpolate the \emph{atomic
  energy} from DFT calculations, and the total energy of an atomic
structure $\sigma$ is given as the sum of the atomic energies of all
atoms
\begin{align}
  E(\sigma)
  = \sum_{i}^{N_{\textup{atoms}}} E_{\textup{atom}}(\sigma_{i}^{R_{c}})
  \label{eq:E_atom}
  \quad .
\end{align}
In equation \eqref{eq:E_atom},
$\sigma_{i}^{R_{c}}=\{\vec{R}_{j},t_{j}:
||\vec{R}_{j}-\vec{R}_{i}||\leq{}R_{c}\}$ captures the local atomic
environment of atom $i$, i.e., the set of the coordinates $\vec{R}_{j}$
and chemical species $t_{j}$ of atoms within a radial cutoff $R_{c}$
from atom $i$.
Note that the number of atoms in $\sigma_{i}^{R_{c}}$ (i.e., within the
cutoff radius) depends directly on the density of the structure, and the
number of chemical species additionally depends on the chemical
composition.

For the construction of a transferable ANN potential, a constant-size
descriptor of the local atomic environment is needed (see
section~\ref{sec:methods-descriptor}).
With such a descriptor of the local atomic environment,
$\widetilde{\sigma}_{i}^{R_{c}}$, the total energy of the ANN potential
is then given by
\begin{align}
  E(\sigma)
  = \sum_{i}^{N_{\textup{atoms}}} \textup{ANN}_{t_{i}}(\widetilde{\sigma}_{i}^{R_{c}})
  \label{eq:ANN-potential}
  \quad ,
\end{align}
where $\textup{ANN}_{t_{i}}$ is the atomic energy ANN potential for
chemical species $t_{i}$.

In the present work, an ANN architecture with two hidden layers each
consisting of 15~nodes was used, giving a descriptor dimension of~44
(see section~\ref{sec:methods-descriptor}) and a total of 931~ANN
parameters.
We previously confirmed that this architecture provides sufficient
flexibility to fit high-dimensional first principles
data.\cite{prB85-2012-045439, pssb250-2013-1191, nl14-2014-2670}

Before the ANN potential training, ten percent of all reference data
points were randomly selected as an independent test set for
cross-validation and were not considered during training.
The ANN potentials were trained using the limited-memory
Broyden-Fletcher-Goldfarb-Shanno (L-BFGS) method~\cite{siam16-1995-1190,
  toms23-1997-550} as implemented in the \emph{atomic energy network}
(ænet) package.\cite{cms114-2016-135}
Training was repeated ten times using different randomly initialized
fitting parameters, and out of this set the optimal ANN fit was
selected.

\subsubsection{Descriptor of the local atomic environment}
\label{sec:methods-descriptor}

The atomic coordinates cannot directly be used as ANN input, as the
number of atoms within the interaction range $R_{c}$ varies with the
structure and density of the material.
Hence, a descriptor of the local atomic environment with constant
dimension is required as the input layer of the atomic energy ANN model
in Eq.\eqref{eq:ANN-potential}.
The descriptor additionally has to obey the invariants of the atomic
energy, i.e., it has to be invariant with respect to rotation,
translation, and the exchange of equivalent atoms.
Several atomic structure descriptors for machine-learning models have
been proposed in the literature.\cite{jcp134-2011-074106,
  jcp139-2013-184118, prb87-2013-184115, prb89-2014-205118,
  ijqc115-2015-1094, ijqc115-2015-1084, arxiv-1704.06439}

Here we employ a recently developed descriptor based on the expansion of
the radial and angular distribution functions (RDF and ADF) that is
numerically efficient and has the advantage that its complexity does not
increase with the number of chemical species.\cite{prb96-2017-14112}
A complete dervation and benchmarks of the method can be found in the
original reference~\citenum{prb96-2017-14112}.
In brief, the RDF and ADF of the local structural environment of atom
$i$ within a cutoff radius $R_{c}$ are defined as
\begin{align}
  \mathrm{RDF}_{i}(r)
  &= \sum_{\mathclap{\;\;\mathbf{R}_{j}\in\,\sigma_{i}^{R_{\textup{c}}}}}
    \delta(r - R_{ij}) \,f_{\textup{c}}(R_{ij})\, w_{t_{j}}
  \label{eq:RDF}
\\
  \mathrm{ADF}_{i}(\theta)
  &= \sum_{\mathclap{\;\;\mathbf{R}_{j},\mathbf{R}_{k}\in\,\sigma_{i}^{R_{\textup{c}}}}}
    \delta(\theta - \theta_{ijk})\,f_{\textup{c}}(R_{ij})\,f_{\textup{c}}(R_{ik})
    \,w_{t_{j}}w_{t_{k}}
  \quad ,
  \label{eq:ADF}
\end{align}
where $R_{ij}$ is the bond distance between atoms $i$ and $j$,
$\theta_{ijk}$ is the bond angle for atoms $i$, $j$, and $k$, and
\begin{align}
  f_{\textup{c}}(r)
  = \begin{cases}
    \frac{1}{2} \Bigl[
      \cos\Bigl(r\cdot\frac{\pi}{R_{\textup{c}}}\Bigr) + 1
    \Bigr]
    & \text{for} \quad r \leq{} R_{c} \\
    0 & \text{else}
  \end{cases}
\end{align}
is a cutoff function.
The choice of the weights $w_{t_{i}}$ for chemical species $t_{i}$ will
be discussed shortly.
Both RDF and ADF automatically obey the invariants of the atomic energy.
Expanding the RDF and ADF in an orthonormal basis set, up to a specified
order $N$, gives an approximate constant-size representation that can be
used as a descriptor for a machine-learning model.

With a basis set of Chebyshev polynomials $\{T_{\alpha}\}$, the
following expressions are obtained for the expansion coefficients of the
RDF
\begin{align}
  \begin{aligned}
  c^{(2)}_{\alpha}
  &= \sum_{\mathclap{\;\;\mathbf{R}_{j}\in\,\sigma_{i}^{R_{\textup{c}}}}}
     T_{\alpha}\Bigl(\frac{2 R_{ij}}{R_{\textup{c}}}-1\Bigr)
     \,f_{\textup{c}}(R_{ij}) \,w_{t_{j}}
    \\
  &\hspace{20ex}\text{with}\quad
    0 \leq R_{ij} \leq R_{c}
  \end{aligned}
  \label{eq:coeff-(2)}
\intertext{and the ADF}
  \begin{aligned}
  c^{(3)}_{\alpha}
  &= \sum_{\mathclap{\;\;\mathbf{R}_{j},\mathbf{R}_{k}\in\,\sigma_{i}^{R_{\textup{c}}}}}
     T_{\alpha}\Bigl(\frac{2 \theta_{ijk}}{\pi}-1\Bigr)
     \,f_{\textup{c}}(R_{ij})\,f_{\textup{c}}(R_{ik})
    \,w_{t_{j}}w_{t_{k}}
    \\
  &\hspace{20ex}\text{with}\quad
    0 \leq \theta_{ijk} \leq \pi
  \label{eq:coeff-(3)}
  \quad .
  \end{aligned}
\end{align}
Note that the arguments of $T_{\alpha}$ in Eqs.~\eqref{eq:coeff-(2)}
and~\eqref{eq:coeff-(3)} are scaled to the interval $[-1.0,+1.0]$ for
which the Chebyshev polynomials are orthogonal with
\begin{equation}
\begin{aligned}
  \int_{-1}^{1}\!\!\! T_{n}(x)T_{m}(x) \frac{\mathrm{d}x}{\sqrt{1 - x^{2}}}
  = \begin{cases}
    \pi           & n=m=0 \\
    \frac{\pi}{2} & n=m\neq{}0 \\
    0             & \text{else}
  \end{cases}
  \\[0.25\baselineskip]
  \quad\text{for}\quad
  x\in{}[-1.0,+1.0]
  \quad .
\end{aligned}
\label{eq:Chebyshev-orthogonality}
\end{equation}

To construct a descriptor for both the local atomic structure
$\{\vec{R}_{i}\}$ and the chemical species $\{t_{i}\}$, two sets of
expansion coefficients are used:
The first set only describes the local structure, and all species
weights $w_{t_{i}}$ in Eqs.~\eqref{eq:E_atom}
through~\eqref{eq:coeff-(3)} are taken to be equal to 1, i.e., all
atomic species are considered to be equivalent in the expansion.
Expanding the RDF and ADF with this choice of $w_{t_{i}}$ yields the
coefficients
$\{{}^{\textup{s}}c_{\alpha}^{(2)}, {}^{\textup{s}}c_{\alpha}^{(3)}\}$
which describe only \emph{structural} features.
The second set of coefficients is obtained by assigning a different
value $w_{t_{i}}$ to each chemical species $t_{i}$, yielding the
expansion coefficients
$\{{}^{\textup{t}}c_{\alpha}^{(2)}, {}^{\textup{t}}c_{\alpha}^{(3)}\}$
that describe the atom \emph{types}.
The combined set
$\{{}^{\textup{s}}c_{\alpha}^{(2)}, {}^{\textup{s}}c_{\alpha}^{(3)},
{}^{\textup{t}}c_{\alpha}^{(2)}, {}^{\textup{t}}c_{\alpha}^{(3)}\}$ is
used as the descriptor for the ANN potentials.
One possible choice of unique species weights for compositions with
$N_{t}$ chemical species is
$w_{t_{i}}=(0,) \pm{}1, \pm{}2, \ldots, \lfloor{}\frac{N_{t}}{2}\rfloor$
where $0$ is only included for odd numbers of
species.~\cite{prb96-2017-14112}

For the present work, we choose $w_{\textup{Li}}$\,=\,-1 and
$w_{\textup{Si}}$\,=\,+1, and $N=11$ Chebyshev polynomials are used for
both the radial and angular expansions.
Hence, the combined descriptor with radial and angular coefficients for
structure and atomic species has a total dimension of $4\times{}11=44$.

\section{Genetic algorithm sampling with a specialized ANN
  potential}
\label{sec:methods-GA}

The stable ground state phases of the \ce{LiSi} alloy and their
crystalline structures, \ce{c-Li8Si8},\cite{jssc173-2003-251}
\ce{c-Li12Si7},\cite{cb119-1986-3576} \ce{c-Li7Si3},\cite{zm71-1980-357}
\ce{c-Li13Si4},\cite{zn30-2014-10} and
\ce{c-Li21Si5},\cite{jssc70-1987-48} are well
known.\cite{jssc37-1981-271, jac496-2010-25}
However, during lithiation and delithiation at room temperature these
crystalline phases are not observed.
Instead, metastable amorphous a-\ce{Li_xSi} structures form that
crystallize into the metastable c-\ce{Li15Si4} phase\cite{am51-2003-1103,
jes151-2004-838, esl7-2004-93, jes154-2007-156, am25-2013-4966}
when the Li potential drops to $\sim$50~mV vs.~\ce{Li+/Li}

To generate realistic amorphous \ce{LiSi} structures as they occur in an
actual Li-ion battery anode, we directly simulate the amorphization
during electrochemical delithiation.
This means, Li is computationally extracted from the fully lithiated and
crystalline \ce{Li15Si4} structure and the atomic positions and the cell
parameters are relaxed at intermediate compositions.

\enlargethispage{\baselineskip}
\subsection{A specialized ANN potential for the sampling of
  near-ground-state \ce{Li_xSi} structures}
\label{sec:specialized-ANN-pot}

To determine the structure of the metastable amorphous \ce{Li_xSi} phase
at different compositions, we employ an evolutionary (or genetic)
algorithm for the sampling of Li/vacancy orderings coupled with a
\emph{specialized} ANN potential.

To train a suitable ANN potential, an initial set of \ce{Li_xSi}
reference structures was generated by isotropic scaling of the lattice
parameters by up to $\pm{}5\%$ and by distortion of the crystalline
c-\ce{Li_xSi} phases, structures of which were obtained from the
Materials Project database.\cite{aplm1-2013-011002}
In addition to the ideal crystal structures, structures with random
Li and Si vacancies were also included.
The resulting initial reference set comprised 725 \ce{Li_xSi} structures
and was used to train a specialized ANN potential for the GA sampling.

We consider this potential to be specialized because it is trained only
to DFT reference calculations of structures that are related to the
crystalline \ce{Li_xSi} structures.
As such, it cannot be expected that the resulting potential would
reproduce the correct energetics of structural motifs that are very
different from those of the crystalline structures, such as different
atomic coordinations and much shorter bond length, as they would occur
at very high energies.
However, we hypothesize that the specialized potential is suitable for
the specific purpose of sampling near-ground-state Li/vacancy
arrangements in delithiated amorphous \ce{Li_{15-x}Si4} structures.

\subsection{Sampling amorphous \ce{Li_xSi} structures with a genetic
  algorithm}
\label{sec:GA-sampling}

\begin{figure*}[tb]
  \centering
  \includegraphics[width=\textwidth]{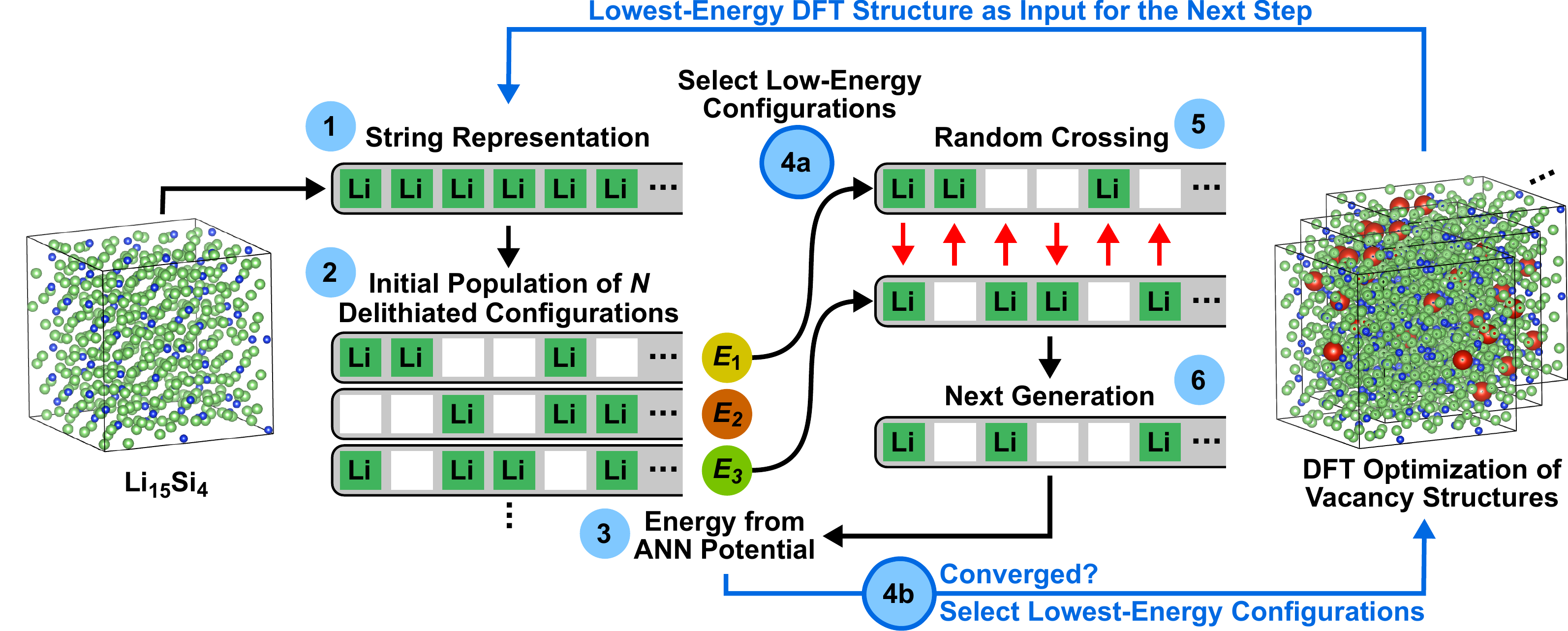}
  \caption{\label{fig:schematic}%
    Schematic of the genetic algorithm (GA) sampling approach using the
    specialized ANN potential.  The GA is used to identify the most
    likely Li atoms to be extracted at each delithiation step, starting
    with the crystalline \ce{Li15Si4} structure.  The atomic positions
    and cell parameters of configurations with intermediate compositions
    are subsequently optimized using DFT.  The magnitude of the energies
    shown in the schematic are $E_{3}<E_{1}<E_{2}$.  Li atoms are shown
    as green balls, Si is blue, and vacancies are red.  See text for
    details of the algorithm.}
\end{figure*}

Genetic algorithms (GAs) are standard global optimization
techniques~\cite{Goldberg1989} and have been routinely applied to atomic
structure optimization problems.\cite{prl75-1995-288,
  jcp124-2006-244704, prb73-2006-224104, prb95-2017-14114}
Here, we are dealing with a simplified optimization problem in that we
only seek to identify those Li atoms that are most likely to be removed
at each delithiation step.
The specific delithiation algorithm employed in the present work is as
follows:
\textbf{(i)}~The GA is used to determine the most stable Li/vacancy
configuration for a supercell of the \ce{Li_{15-x}Si_{4}} structure.
\textbf{(ii)}~The atomic coordinates and lattice parameters of at
least the~30 most stable configurations as predicted by the GA are optimized
with DFT.
\textbf{(iii)}~The most stable (lowest-energy) structure as determined
by DFT is used as the starting point for the next delithiation step, and
the scheme is continued with step~\textbf{(i)}.

A schematic of the GA method used in the present work is shown in
\textbf{Fig.~\ref{fig:schematic}}.
In detail, the GA involves the following individual steps:
\begin{enumerate}
\item[0.] Input for the GA is the structure of one particular
  \ce{Li_xSi} composition.  For the first delithiation step, a supercell
  of the ideal \ce{Li15Si4} structure is used.
\item[1.] The atomic configuration is represented as a vector (or
  string) in which only Li sites are considered;
\item[2.] An initial population of $N$ trial configurations is generated
  by randomly removing Li atoms from the input structure (delithiation)
  to realize the specific \ce{Li_{x$_1$}Si} composition of the present
  delithiation step;
\item[3.] The energy of each trial configuration is evaluated using the
  ANN potential.  If the optimization has converged and no lower energy
  was determined over a certain number of steps the algorithm is
  continued with step 4b, else the optimization is continued with step
  4a;
\item[4a.] For the following steps, each two trial configurations from
  the current population are selected with probabilities that are
  proportional to their energy such that lower energy means higher
  selection probability. This selection method is sometimes called
  \emph{roulette wheel selection};~\cite{Goldberg1989}
\item[5.] $N$ additional trials are generated by combination
  (\emph{crossing}) of two selected trials from the current population.
  Each new trial configuration is further subject to random changes with
  a mutation probability of $p_{m}$ (not shown in
  \textbf{Fig.~\ref{fig:schematic}});
\item[6.] The energies of the new trial configurations are evaluated
  using the ANN potential, and the algorithm continues with step 3;
\item[4b.] Once the GA optimization has converged or a set number of
  steps have been completed, the $M$ configurations with the lowest
  energies are prepared for subsequent geometry optimizations with
  DFT.
\end{enumerate}
For the present work, we used a population size of $N$\,=\,32 trials and
a mutation rate of $p_{m}$\,=\,10\%.
At least $M=30$ configurations were optimized with DFT at each
composition.
A Python implementation of the above GA algorithm can be obtained from
\href{http://ga.ann.atomistic.net}{http://ga.ann.atomistic.net}.

Using the GA approach described above coupled with the specialized ANN
potential of section~\ref{sec:specialized-ANN-pot}, two different
supercells of the c-\ce{Li15Si4} phase with compositions \ce{Li60Si16}
(76~atoms) and \ce{Li480Si128} (608~atoms) were delithiated.
The small \ce{Li60Si16} cell was delithiated in intervals of each 2~Li
atoms, and the large \ce{Li480Si128} cell was delithiated in steps of
8~Li atoms.

\begin{figure*}[tb]
  \centering
  \includegraphics[width=0.7\textwidth]{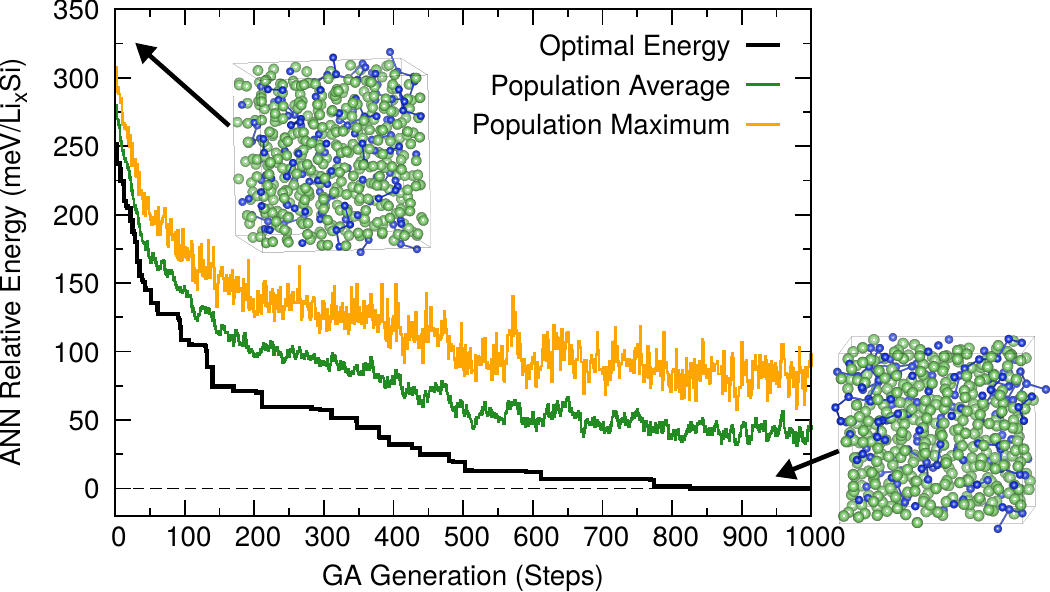}
  \caption{\label{fig:GA-energies}%
    Energy of the \ce{Li360Si128} configurations during optimization
    with the genetic algorithm (GA) coupled with the specialized ANN
    potential.  The energy of the current best configuration is shown as
    black line (optimal energy), the green line corresponds to the
    average energy of the GA population of 32~trials, and the yellow
    line is the current maximal energy.  The insets show the initial
    structure and the final structure with lowest energy after
    optimization with DFT.  Li and Si atoms are colored green and blue,
    respectively.}
\end{figure*}
As one concrete example, the course of the energy during GA optimization
of an atomic configuration with composition \ce{Li360Si128} is shown in
\textbf{Fig.~\ref{fig:GA-energies}}.
As seen in the figure, after around 500~GA steps the energy only changes
by around 13~meV/\ce{Li_xSi}, indicating that the optimization has
reasonably converged.

A total of 1,263 structures from this GA sampling were selected for
subsequent DFT evaluation and geometry optimization, and together with
the initial reference set they form the basis for the first principles
phase diagram discussed below.
%
\vspace*{\fill}

\section{Molecular dynamics sampling with a general ANN potential}
\label{sec:MD-sampling}

The GA methodology described above makes two approximations that may
intuitively not seem justified:
(i)~The GA sampling does not consider structural relaxations (though,
the final 30 or more low-energy configurations are fully optimized), and
(ii)~the ANN potential is specialized for the GA sampling and would not
be suitable for other applications.
To verify that the GA sampling generated genuinely low-energy metastable
amorphous structures, we compare the resulting phase diagram with the
one obtained from heat-quench molecular dynamics (MD) simulations.

All MD simulations were carried out using the Tinker software
package~\cite{jocc8-1987-1016} and a Parrinello-Bussi
thermostat~\cite{jcp126-2007-14101} in the $NVT$ ensemble.
Generally, a time step of 2~fs was used for the integration of the
equation of motion with the Verlet algorithm.\cite{pr159-67-98.pdf}

\subsection{ANN potential construction and molecular dynamics
  simulations}
\label{sec:general-ANN-pot}

To carry out reliable MD simulations, a fully \emph{general} ANN
potential is required.
For the training of such an ANN potential, a more extensive set of DFT
reference calculations is needed that also includes local structural
motifs that do not occur in near-ground-state bulk structures.
This means, also unphysical bonding situations and lattice parameters as
well as unusual coordinations should be present in the reference data
set, so that structures that exhibit those features are not artificially
overstabilized during MD simulations.
Therefore, we also included clusters with up to $\approx$200~atoms, and
surface slab structures that were truncated from the bulk in addition to
further bulk reference structures.

\begin{figure*}[tb]
  \centering
  \includegraphics[width=0.8\textwidth]{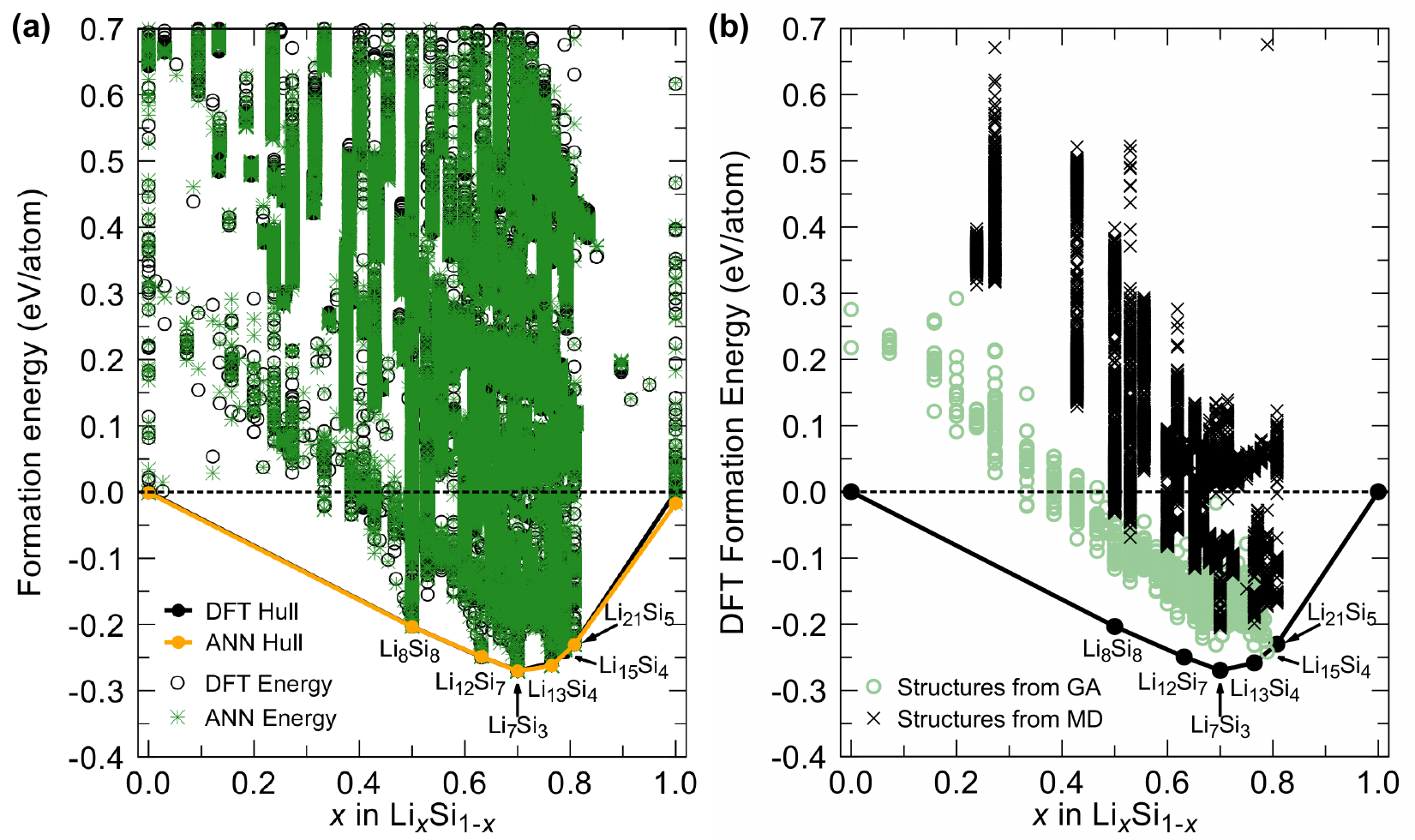}
  \caption{\label{fig:GA-vs-MD}%
    \textbf{(a)}~Phase diagram based on the formation energies of all
    $\sim$45,000 \ce{Li_xSi} structures including bulk, surface slab,
    and cluster structures from both training and test set.  The
    energies predicted by the general ANN potential are shown as green
    stars and the DFT reference energies are black circles.
    \textbf{(b)}~Only DFT formation energies of those structures sampled
    by the GA with the specialized ANN potential (green circles) and
    only those generated during MD simulations with the general ANN
    potential (black crosses).}
\end{figure*}
The additional structures were generated by repeated MD simulations.
In addition to the crystalline \ce{Li_xSi} structures, the lowest
energy structures from the GA sampling of section~\ref{sec:GA-sampling}
were also used as starting points for MD simulations.
Hence, structures with up to 608~atoms were considered.

Short (<\,10~ps) ANN potential MD simulations of the input \ce{Li_xSi}
structures at very high temperatures up to 3,000~K were employed to
amorphize the structures.
These heating simulations were followed by subsequent 2~ns long
simulations at lower temperatures (between 400~K and 1200~K) to obtain
equilibrated low-energy structures.
400~structures (every 5~ps) along each 2~ns long MD trajectory were
recomputed with DFT single-point calculations and subsequently included
in the reference data set.

MD simulations and ANN potential re-training were repeated until a low
root mean squared error (RMSE) relative to the DFT energies was obtained
and all \ce{Li_xSi} ground states of the ANN potential and DFT energies
were in agreement.
Iterative ANN potential training based on MD simulations is also
described in more detail in reference~\citenum{cms114-2016-135}.

In total, around 45,000~reference structures were used for the training
of the general ANN potential, including the reference structures of the
specialized potential, the structures from GA sampling, and the
additionally generated bulk, slab, and cluster structures.
Ten percent of this reference data set, around 4,500~randomly selected
structures, were set aside as an independent test set for cross
validation and were not used in the final ANN potential training.
Simultaneously, 10 different ANN potentials were trained with different
random initial ANN weight parameters on the remaining 40,500 data
points. Out of these potentials, the one with the smallest overall error
relative to the DFT reference energies reproduces the DFT ground state
phase diagram most accurately and was therefore selected for the
subsequent analysis.

The selected ANN potential achieves an RMSE of 7.7~meV/atom and a mean
absolute error (MAE) of 5.9~meV/atom for the test set.
The RMSE and MAE for the training set are 6.3~and 5.7~meV/atom,
respectively.

\textbf{Figure~\ref{fig:GA-vs-MD}a} shows a comparison of the formation
energies predicted by the ANN potential and their DFT references for all
structures in the reference data set.
As seen in the figure, all features of the DFT formation energies are
correctly reproduced by the ANN potential, and the energies of the
individual data points are in good agreement.
In the present work, this general ANN potential is only used for the
purpose of validating the results of the approximate GA sampling.
However, this general potential will enable the investigation of other
properties of the LiSi alloy in future projects.

\begin{figure*}[tb]
  \centering
  \includegraphics[width=0.9\textwidth]{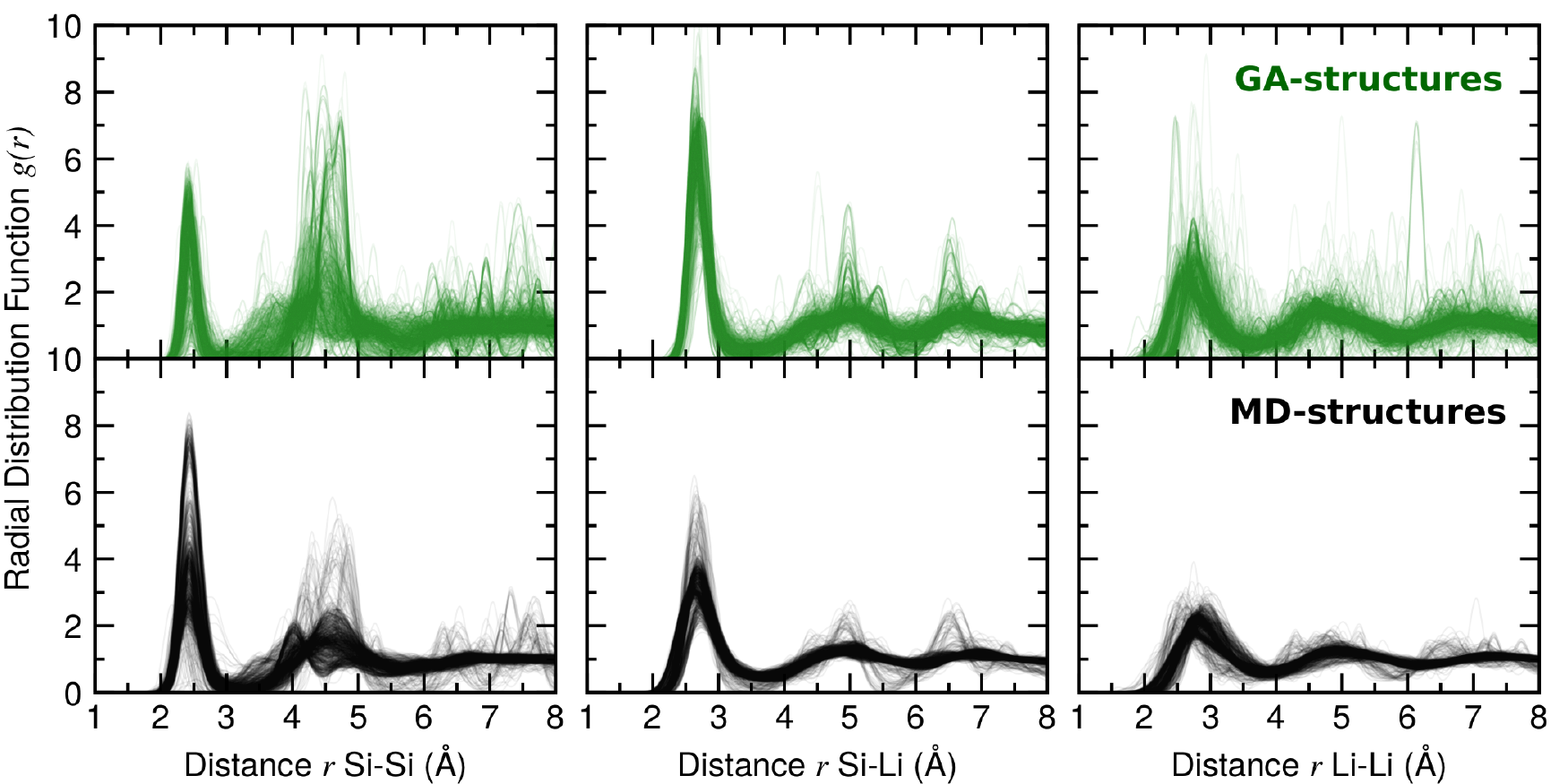}
  \caption{\label{fig:RDF-GA-MD}%
    Superposition of the radial distribution functions (RDF) of
    each 1,000 randomly selected structures obtained from the GA
    sampling (top) and from the MD sampling (bottom).}
\end{figure*}
\subsection{Comparison of the amorphous phase space sampled by GA and MD}
\label{sec:phase-space}

Having at hand the two sets of independently sampled amorphous
\ce{Li_xSi} structures from the approximate GA sampling
(section~\ref{sec:GA-sampling}) and from the extensive MD heat-quench
sampling (section~\ref{sec:general-ANN-pot}), we can now directly assess
the differences.
\textbf{Figure~\ref{fig:GA-vs-MD}b} shows only those DFT formation
energies belonging to structures obtained from the GA sampling (light
green circles) and from the MD sampling (black crosses).
Clearly, the GA sampling yielded consistently lower energy structures
than the MD sampling at all considered \ce{Li_xSi} compositions.
Despite the approximations made in the GA sampling, the sampled
structures are within a small energy range of around 100~meV/atom above
the lowest energy of all a-\ce{Li_xSi} for each composition.
Thus, the ANN-potential-assisted GA sampling is successful in
determining structure models of low-energy metastable amorphous phases,
and a fully converged and general ANN potential is not required if that
is the only objective.

\section{Discussion}
\label{sec:discussion}

\begin{figure*}
  \centering
  \includegraphics[width=0.9\textwidth]{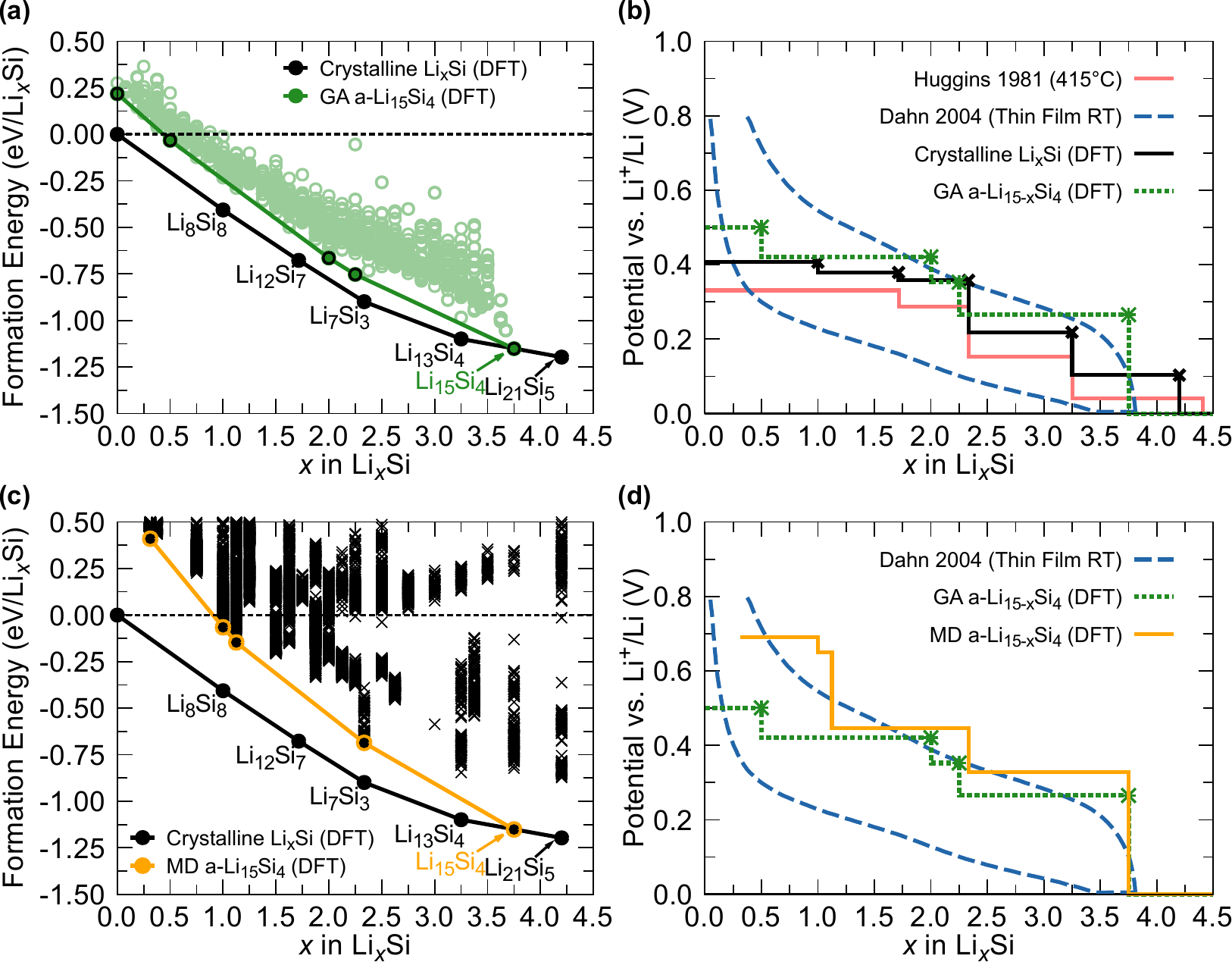}
  \caption{\label{fig:LiSi}%
    \textbf{(a)}~DFT formation energies of the amorphous
    \ce{Li_{480-x}Si128} structures generated by the GA sampling (green)
    along with the convex hull of the crystalline LiSi phases.  In this
    figure, the composition axis has been normalized to a constant Si
    content (\ce{Li_xSi}) which reflects the Li storage capacity per Si
    atom in a Si anode.  \textbf{(b)}~Computed voltages corresponding to
    the lower convex hulls of the formation energies in \textbf{(a)}
    compared to measured voltage profiles from the literature.  The DFT
    voltage profile based on the crystalline \ce{Li_xSi} phases (black
    solid line) can be compared to the equilibrium voltages measured by
    Huggins~et~al.\cite{jssc37-1981-271} (light red line), while the DFT
    voltages of the amorphous phases (dotted green line) more closely
    correspond to the voltage profile by Dahn et
    al.\cite{jes151-2004-838} (dashed blue line) measured for a thin
    film amorphous Si anode.  \textbf{(c)}~DFT formation energies of the
    structures from MD sampling.  \textbf{(d)}~Comparison of the
    computed amorphous \ce{Li_xSi} voltage profiles based on the GA and
    MD structures.}
\end{figure*}

The coupling of a GA with a specialized MLP (GA-MLP) as proposed in this
work is designed to generate low-energy metastable amorphous structures,
and we demonstrated that this is accomplished.
However, the comparison of the GA and MD sampling results in the
previous section show that the sampling strategy has a pronounced effect
on the structures and energetics.
In our simulations, computationally quenching a melt gives rise to
energetically quite different results than computational delithiation.
It should be noted that the outcome of the heat-quench simulations
depends on the cooling rate of the quenching step and on the simulated
time scale, as rapid quenching can trap the system in high-energy
states.
However, the 2~ns long MD simulations of the present work are already
far beyond the time scales that can be reached with first principles
methods.
In practice, it depends on the concrete application which sampling
strategy is most appropriate to model an amorphous phase.
When modeling glasses, for example, heat-quench simulations may be most
appropriate.
In the present example of electrochemical amorphization, simulated
delithiation more closely resembles the actual mechanism.

To better understand the differences between the structures generated by
the two different sampling approaches, we evaluated the radial pair
distribution functions (RDFs) for each 1,000 structures from the GA and
MD sampling. A superposition of the RDFs is shown in
\textbf{Fig.~\ref{fig:RDF-GA-MD}}. As seen in the figure, the majority
of structures generated by the MD heat-quench simulations exhibit little
correlations beyond the first coordination shell, and at a distance of 8
\AA{} the RDFs approach a value of 1 indicating absence of ordering. In
contrast, the structures generated using the GA sampling approach
exhibit stronger long-range correlations reminiscent of the ground-state
crystal structures as can be seen by comparison with the RDFs reported
by Dahn et al. for the crystalline LiSi phases.\cite{jes156-2009-454}
The reason for this difference is likely that the heat-quench
simulations start at high temperatures at which no long-range ordering
is present, and the cooling-rate is apparently too fast to allow for the
emergence of ordering. The GA sampling, on the other hand, simulates the
amorphization by lithium extraction without temperature effects.

One concrete example related to the \ce{Li_xSi} alloy is the voltage
relative to \ce{Li+/Li}.
Computationally, the average conversion voltage between two \ce{Li_xSi}
phases at $T$\,=\,0\,K can be well approximated based on the
first-principles energy differences relative to Li
metal~\cite{jps68-1997-664, ncm2-2016-16002}
\begin{equation}
\begin{aligned}
  \overline{V} =
  -\frac{E(\ce{Li_{x_{1}}Si}) - E(\ce{Li_{x_{2}}Si}) - (x_{1}-x_{2})E(\ce{Li})}{(x_{1}-x_{2})F}
  \\[0.25\baselineskip]
  \quad\text{with}\quad x_{1} > x_{2}
  \quad ,
\end{aligned}
\label{eq:voltage}
\end{equation}
where $F$ is Faraday's constant.
In Eq.~\ref{eq:voltage}, the energies E\ce{(Li_xSi_y)} are obtained from
the lower convex hull of the formation energy which corresponds to the
T = 0 K phase diagram. At higher temperatures, the steps of the voltage
profile will be smoothed out to some extend.\cite{jelecom6-2004-1045}
\textbf{Figure~\ref{fig:LiSi}a} shows the formation energies of the
structures from GA sampling for a lithium content that is normalized to
one Si atom, so that the $x$-axis corresponds directly to the capacity
in battery applications.
The corresponding computational 0\,K voltage profiles for the
crystalline and amorphous phases are shown in
\textbf{Fig.~\ref{fig:LiSi}b} along with experimentally measured voltage
profiles taken from the literature.\cite{jssc37-1981-271,
  jes151-2004-838}
Note that the T = 0 K voltage profile of the crystalline LiSi phases
(solid black line in Fig. 5b) agrees well with the measured
T = $415^\circ$C voltage profile showing that the 0 K voltage profile is
a reasonable approximation.
The formation energies and voltage profile from MD sampling are
equivalently visualized in \textbf{Fig.~\ref{fig:LiSi}c} and
\textbf{Fig.~\ref{fig:LiSi}d}.
Evidently, the GA-generated formation energies shown in
\textbf{Fig.~\ref{fig:LiSi}a} give rise to a qualitatively and
quantitatively different 0~K phase diagram as those from the MD sampling
(\textbf{Fig.~\ref{fig:LiSi}c}), as defined by the lower convex hull of
the formation energies.
The higher energies of the structures from MD sampling for low Li
contents and the overall greater slope of the convex hull result in a
much steeper voltage profile than the GA counterpart, as seen in
\textbf{Fig.~\ref{fig:LiSi}d}.
The large hysteresis in the experimental voltage profile prevents a
quantitative comparison with the computed voltages, though the voltage
profile based on the energetics from GA sampling is closer to the
experimental average.

Finally, we stress that machine-learning-assisted sampling is neither
limited to the present application (delithiation) nor to genetic
algorithms.
The main message of the present work is instead that a specialized
machine-learning potential is sufficient for structural sampling in a
limited domain.
Other amorphization mechanisms, such as amorphization caused by
off-stoichiometries (e.g., from doping) can be modeled with an
equivalent sampling setup.
As an alternative to a GA for the sampling of low-energy configurations,
we could have also employed other techniques such as Monte-Carlo
simulations (i.e., simulated annealing).\cite{nl14-2014-2670}
It should be noted that the GA sampling does not describe real-time
dynamics and the present methodology assumes quasi-equilibration at each
composition. The obtained voltages therefore correspond to equilibrium
voltages, i.e., slow delithiation. In contrast, the MD sampling
corresponds to a delithiation rate that is determined by the length of
the simulated trajectories between delithiation steps (2 ns) which might
provide an explanation of the higher voltages obtained from the MD
sampling.

\section{Conclusions}
\label{sec:conclusions}

Using the example of the amorphous LiSi alloy, we showed how specialized
machine-learning potentials can be used to speed up the first-principles
sampling of complex structure spaces.
Our methodology is based on a combination of a genetic algorithm (GA)
with an artificial neural network (ANN) potential.
We demonstrated that this ANN-assisted sampling is successful in
determining low-energy amorphous structures and is computationally more
efficient than the construction of a converged general ANN potential.
Using molecular dynamics heat-quench simulations, we confirmed that the
metastable structures generated by ANN-assisted GA sampling are
consistently within a low energy range.
The herein described method is not limited to a specific material or
amorphization mechanism but is generally applicable to the modeling of
amorphous and disordered materials.


\section{Acknowledgments}

The authors thank China Automotive Battery Research Institute Co., Ltd.\
and General Research Institute for NonFerrous Metals (GRINM) for
financial support.
This work used the computational facilities of the Extreme Science and
Engineering Discovery Environment (XSEDE), which is supported by
National Science Foundation grant no.~ACI-1053575.
Additional computational resources from the University of California
Berkeley, HPC Cluster (SAVIO) are also gratefully acknowledged.

\raggedright
\bibliography{Artrith-Preprint-Bibliography}

\end{document}